\documentclass[letter]{aa} 
\usepackage{graphicx}
\usepackage{longtable}
\usepackage{txfonts}
\begin{document}
   \title{The large amplitude outburst of the young star HBC 722 in NGC 7000/IC 5070, a new FU Orionis candidate}


   \author{E. H. Semkov\inst{1}
          \and    
          S. P. Peneva\inst{1}
          \and
          U. Munari\inst{2}
          \and
          A. Milani\inst{3}
          \and
          P. Valisa\inst{3}}

   \offprints{E. Semkov (esemkov@astro.bas.bg)}

    \institute{Institute of Astronomy and National Astronomical Observatory, Bulgarian Academy of Sciences,
              72 Tsarigradsko Shose blvd., BG-1784 Sofia, Bulgaria,
              \email{esemkov@astro.bas.bg}
         \and
             INAF Osservatorio Astronomico di Padova, Sede di Asiago, I-36032 Asiago (VI), Italy
         \and
             ANS Collaboration, Astronomical Observatory, I-36012 Asiago (VI), Italy}

   \date{Received ; accepted }


  \abstract
   {The investigations of the photometric and spectral variability of PMS stars are essential to a better understanding of the early phases of stellar evolution. 
   We are carrying out a photometric monitoring program of some fields of active star formation.  
   One of our targets is the dark cloud region between the bright nebulae NGC 7000 and IC 5070.}
   {We report the discovery of a large amplitude outburst from the young star HBC 722 (LkH$\alpha$ 188 G4) located in the region of NGC 7000/IC 5070.      On the basis of photometric and spectroscopic observations, we argue that this outburst is of the FU Orionis type.}
   {We gathered photometric and spectroscopic observations of the object both in the pre-outburst state and during a phase of increase in its brightness. 
   The photometric $BVRI$ data (Johnson-Cousins system) that we present were collected from April 2009 to September 2010.  
   To facilitate transformation from instrumental measurements to the standard system, fifteen comparison stars in the field of HBC 722 were calibrated in the $BVRI$ bands.  
   Optical spectra of HBC 722 were obtained with the 1.3-m telescope of Skinakas Observatory (Crete, Greece) and the 0.6-m telescope of Schiaparelli Observatory in Varese (Italy).}
   {The pre-outburst photometric and spectroscopic observations of HBC 722 show both low amplitude photometric variations and an emission-line   spectrum typical of T Tau stars.  
   The observed outburst started before May 2010 and reached its maximum brightness in September 2010, with a recorded $\Delta$$V$$\sim$ 4\fm7 amplitude.  
   Simultaneously with the increase in brightness the color indices changed significantly and the star became appreciably bluer.  
   The light curve of HBC 722 during the period of rise in brightness is similar to the light curves of the classical FUors - FU Ori and V1057 Cyg.        The spectral observations during the time of increase in brightness showed significant changes in both the profiles and intensity of the spectral lines.  
   Only H$\alpha$ remained in emission, while the H$\beta$, Na I 5890/5896, Mg I triplet 5174, and Ba II 5854/6497 lines were in strong absorption.}
   {}


   \titlerunning{The large amplitude outburst of the young star HBC 722 in NGC 7000/IC 5070}
   \maketitle
%

\section{Introduction}

Photometric variability is a fundamental characteristic of pre-main sequence (PMS) stars.  
One of the most interesting phenomena of early stellar evolution is the large increase in brightness (usually termed {\em an outburst}) that has been recorded for some PMS stars. 
These outbursts can be grouped into two main types, named after their respective prototypes: FU Orionis (FUor; Ambartsumian 1971) and EX Lupi (EXor; Herbig 1989).  
Both type of stars seem to be related to low-mass PMS objects (T Tauri stars) with massive circumstellar disks, and their outbursts are generally attributed to infall into the central star of material from a circumstellar disk (Hartmann \& Kenyon 1985).

Only a few FUor objects are known, which all share the same defining characteristics: a $\Delta$$V$$\approx$4-5 mag outburst amplitude, an F-G
supergiant spectrum during outbursts, association with reflection nebulae, location in star-forming regions, a strong LiI~6707~\AA\ in absorption,
H$\alpha$ and Na I 5890, 5896 \AA\ displaying P-Cyg profiles, and CO bands in near-infrared spectra (Herbig 1977; Reipurth 1990).  
The outbursts of FUor objects last for several decades, and the rise time is faster than the decline.

EXor objects display frequent, irregular and relatively brief outbursts (lasting from a few months to a few years) of several magnitudes amplitude ($\Delta$$V$$\approx$3-5).  
During these events, the cool spectrum of the quiescence is veiled, and strong emission lines from single ionized metals are observed together with  reversed P-Cyg absorption components (Herbig 2007; Aspin et al. 2010).

During an optical photometric monitoring of star-forming regions, we discovered in August 2010 a large amplitude outburst from a T Tauri
star located in the dark clouds (so-called "Gulf of Mexico") between NGC 7000 (the North America Nebula) and IC 5070 (the Pelican Nebula). 
The announcement of the outburst discovery was made by Semkov \& Peneva (2010a), and Munari et al. (2010) provided a preliminary report of the
spectroscopic characteristics of the outburst at optical wavelengths. 
Semkov \& Peneva (2010b) and Leoni et al. (2010) reported additional details of infrared and optical photometry of the outburst. 
The star has coordinates R.A. = 20$^{\rm h}$58$^{\rm m}$17\fs03 (J2000) and DEC. = +43\degr 53\arcmin 43\farcs4 (J2000) and is a member of a small group of H$\alpha$ emission objects in the vicinity of LkH$\alpha$ 188, a region characterized by active star formation.  No other outburst has been recorded from this object, for which little is known about the photometric behavior quiescence.  
The only available spectroscopic study of this object was published by Cohen \& Kuhi (1979), who observed the star in 1977 when it was in quiescence.  
They classified the absorption spectrum as K7-M0 and listed the equivalent widths of the H$\alpha$, H$\beta$, and [OI] 6300 spectral lines, the only three lines visible in emission in their spectra.  
Cohen \& Kuhi (1979) designated the star as LkH$\alpha$ 188 G4.  
It is included in the catalogs of emission-line stars compiled by Herbig \& Bell (1988, with designation HBC 722), and by Kohoutek \& Wehmeyer (1999, where it is indicated as [KW97] 53-18).  
It was also recorded by the 2MASS survey as source 20581702+4353433, which was measured at $J$=13.205 $\pm$0.012, $H$=12.136 $\pm$0.052, and $K$=11.412 $\pm$0.016 on 2000 June 10. 

\section{Observations}

We present $\it BVRI$ photometric data collected from April 2009 to September 2010.  
Our photometric observations were performed with the 50/70/172 cm Schmidt telescopes of the National Astronomical Observatory Rozhen (Bulgaria) and the 1.3-m Ritchey-Chretien telescope of the Skinakas Observatory\footnote{Skinakas Observatory is a collaborative project of the University of Crete, the Foundation for Research and Technology - Hellas, and the Max-Planck-Institut f\"{u}r Extraterrestrische Physik.} of the Institute of Astronomy, University of Crete (Greece).  
We used an ANDOR DZ 436 CCD camera with the 1.3-m RC telescope, and a FLI PL 16803 CCD camera with the Schmidt telescope.  
All frames were exposed through a set of standard Johnson-Cousins filters.  
Aperture photometry was performed using DAOPHOT routines.

Using our photometric observations from 2007, 2008, and 2009, a comparison sequence of fifteen stars in the field around HBC 722 was calibrated in
$BVRI$ (Semkov \& Peneva 2010b).  
Standard stars from Landolt (1992) were used as a reference.  
Table 1 contains photometric data for the $BVRI$ comparison sequence. 
The corresponding mean errors in the mean are also listed. 
The stars are labeled from A to O in order of their V-band magnitude.  
In regions of star formation such as NGC 7000, a great percentage of stars can be photometric variables.  
Therefore, some of our standard stars may be low amplitude variables and we advise observers to use our photometric sequence with care.  
The finding chart of the comparison sequence is presented in Fig.  1.  
The field is $8\arcmin\times8\arcmin$, north is at the top and east to the left.  
The chart is retrieved from the STScI Digitized Sky Survey Second Generation Red.

Figure~2 compares two $R$-band CCD images obtained before and during the outburst of HBC 722.  
The emergence of a weak (presumably reflection) nebula around the star is visible in the outburst image.  
To minimize the light from the surrounding nebula and at the same time avoid contamination from adjacent stars, our photometry was obtained through an aperture of 4\arcsec in radius, and the background annulus extended from 13\arcsec to 19\arcsec.
 
\begin{figure}
   \centering
   \includegraphics[width=8cm]{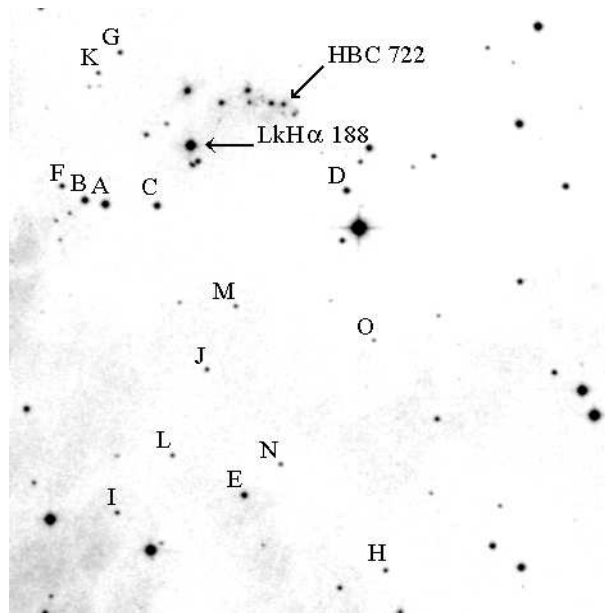}
   \caption{Finding chart for the $BVRI$ comparison sequence around HBC~722}
    \end{figure}

\begin{table*}
\caption{Photometric data for the $BVRI$ comparison sequence.}
\label{table:2}
\centering
\begin{tabular}{ccccccccc}
\hline\hline
\noalign{\smallskip}
Star &  $B$ & $\sigma_B$ & $V$ & $\sigma_V$ & $Rc$ & $\sigma_R$& $Ic$ & $\sigma_I$ \\
\noalign{\smallskip}
\hline
\noalign{\smallskip}
A	& 16.304	& 0.009	& 14.974	& 0.011	& 14.090	& 0.022	& 13.301	& 0.020\\
B	& 16.469	& 0.008	& 15.417	& 0.008	& 14.779	& 0.016	& 14.238	& 0.010\\
C	& 17.270	& 0.018	& 15.823	& 0.014	& 14.882	& 0.021	& 14.001	& 0.021\\
D	& 17.024	& 0.025	& 15.867	& 0.016	& 15.124	& 0.021	& 14.542	& 0.021\\
E	& 17.259	& 0.016	& 16.042	& 0.010	& 15.239	& 0.022	& 14.633	& 0.022\\
F	& 18.636	& 0.070	& 17.246	& 0.016	& 16.322	& 0.030	& 15.581	& 0.032\\
G	& 18.949	& 0.057	& 17.551	& 0.031	& 16.588	& 0.025	& 15.749	& 0.032\\
H	& 19.180	& 0.092	& 17.789	& 0.036	& 16.820	& 0.024	& 16.016	& 0.039\\
I	& 19.401	& 0.145	& 17.978	& 0.037	& 16.984	& 0.043	& 16.093	& 0.072\\
J	& 19.496	& 0.093	& 18.018	& 0.048	& 17.044	& 0.029	& 15.978	& 0.075\\
K	& 20.055	& 0.287	& 18.287	& 0.082	& 17.011	& 0.060	& 15.434	& 0.062\\
L	& 19.884	& 0.239	& 18.375	& 0.057	& 17.291	& 0.030	& 16.351	& 0.079\\
M	& 20.143	& 0.167	& 18.528	& 0.087	& 17.530	& 0.039	& 16.535	& 0.070\\
N	& 20.315	& 0.231	& 18.682	& 0.060	& 17.472	& 0.051	& 16.438	& 0.055\\
O	& 20.581	& 0.164	& 18.921	& 0.065	& 17.912	& 0.039	& 16.772	& 0.102\\
\hline                                   
\end{tabular}
\end{table*}

The results of our photometric observations of HBC 722 are summarized in Table 2.  
The columns provide the JD of observation, $\it IRVB $ magnitudes of HBC 722, and the telescope we used.  
Typical errors in the reported magnitudes are $0\fm01$-$0\fm02$ for $I$ and $R$-band data, $0\fm02$-$0\fm06$ for $V$, and $0\fm03-0\fm10$ for $B$-band.

   \begin{table*}
   \centering
   \small
   \caption[]{$\it BVRI $ photometric observations of HBC 722 for the period Apr. 2009 - Sep. 2010}
   \begin{tabular}{ccccclcccccl}
            \hline \hline
            \noalign{\smallskip}
      JD (24...) & $\it B$  & $\it V$ & $\it R$ & $\it I$  &  Telescope & JD (24...) & $\it B$  & $\it V$ & $\it R$ & $\it I$  &  Telescope\\
            \noalign{\smallskip}
            \hline
            \noalign{\smallskip}
54938.551 &       & 18.12 & 16.80  & 15.24 & Schmidt & 55421.326 & 15.84 & 14.29 & 13.24 & 12.11 & 1.3-m\\
55000.499 & 19.80 & 18.42 & 17.02  & 15.23 & 1.3-m   & 55422.258 & 15.76 & 14.24 & 13.20 & 12.07 & 1.3-m\\
55009.522 & 19.74 & 18.23 & 16.86  & 15.22 & 1.3-m   & 55423.256 & 15.70 & 14.18 & 13.14 & 12.04 & 1.3-m\\
55011.458 &       & 18.32 & 16.99  & 15.31 & Schmidt & 55424.251 & 15.68 & 14.15 & 13.10 & 12.00 & 1.3-m\\
55016.501 & 19.97 & 18.33 & 16.93  & 15.28 & 1.3-m   & 55425.250 & 15.61 & 14.07 & 13.04 & 11.94 & 1.3-m\\
55022.502 & 19.82 & 18.42 & 17.01  & 15.24 & 1.3-m   & 55426.599 & 15.54 & 14.03 & 12.99 & 11.89 & 1.3-m\\
55027.412 &       & 18.40 & 17.10  & 15.33 & Schmidt & 55427.591 & 15.54 & 14.02 & 12.97 & 11.87 & 1.3-m\\
55028.380 &       &       & 16.91  & 15.19 & Schmidt & 55428.590 & 15.57 & 14.05 & 13.00 & 11.88 & 1.3-m\\
55035.502 & 19.80 & 18.18 & 16.82  & 15.18 & 1.3-m   & 55429.589 & 15.58 & 14.05 & 12.99 & 11.88 & 1.3-m\\
55044.338 & 19.90 & 18.33 & 16.98  & 15.28 & 1.3-m   & 55432.519 & 15.37 & 13.86 & 12.83 & 11.75 & 1.3-m\\
55065.292 &       & 18.35 & 16.88  & 15.17 & Schmidt & 55433.470 & 15.32 & 13.81 & 12.79 & 11.71 & 1.3-m\\
55111.330 &       & 18.45 &        & 15.29 & Schmidt & 55434.290 & 15.34 & 13.83 & 12.81 & 11.72 & 1.3-m\\
55113.243 &       & 18.38 & 17.10  & 15.30 & Schmidt & 55435.310 & 15.36 & 13.84 & 12.83 & 11.74 & 1.3-m\\
55156.197 &       & 18.47 & 17.00  & 15.25 & Schmidt & 55439.237 & 15.33 & 13.82 & 12.78 & 11.70 & 1.3-m\\
55157.227 &       & 18.25 & 16.90  & 15.19 & Schmidt & 55440.238 & 15.28 & 13.77 & 12.75 & 11.68 & 1.3-m\\
55330.436 & 18.96 & 17.37 & 16.04  & 14.60 & Schmidt & 55447.420 & 15.17 & 13.69 & 12.70 & 11.62 & Schmidt\\
55358.424 & 18.89 & 17.26 & 16.07  & 14.53 & Schmidt & 55448.326 & 15.16 & 13.68 & 12.70 & 11.63 & Schmidt\\
55415.466 & 16.10 & 14.59 & 13.53  & 12.38 & Schmidt & 55449.407 & 15.20 & 13.72 & 12.73 & 11.64 & Schmidt\\
55416.433 & 16.04 & 14.51 & 13.45  & 12.32 & Schmidt & 55458.218 & 15.15 & 13.65 & 12.65 & 11.59 & 1.3-m\\
55420.497 & 15.84 & 14.30 & 13.25  & 12.12 & 1.3-m   & 55459.513 & 15.19 & 13.69 & 12.67 & 11.61 & 1.3-m\\
\noalign{\smallskip}
\hline
\end{tabular}
\end{table*}

  \begin{figure}
   \centering
   \includegraphics[width=4cm]{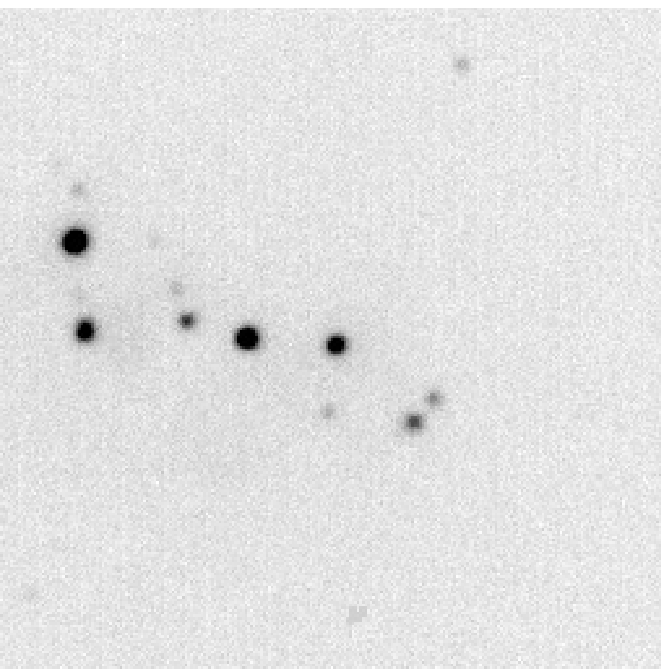}
   \includegraphics[width=4cm]{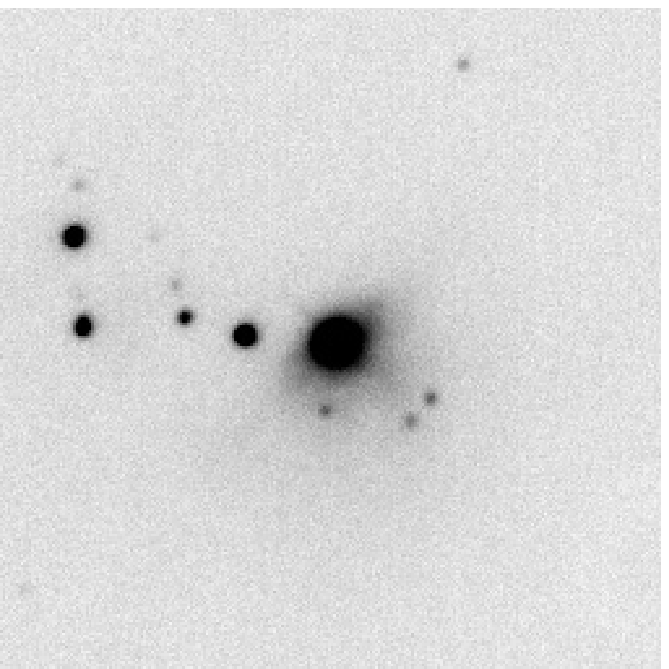}
   \caption{CCD frames of HBC 722 obtained with the Skinakas Observatory
   1.3-m RC telescope through a $R$ filter.  {\it Left}: on 2009 Jul. 31. 
   {\it Right}: on 2010 Aug.  26.  The appearance of small reflection nebula
   around the star is observed on the second frame.}
    \end{figure}
   
  \begin{figure*}
   \centering
   \includegraphics[width=7cm,angle=270]{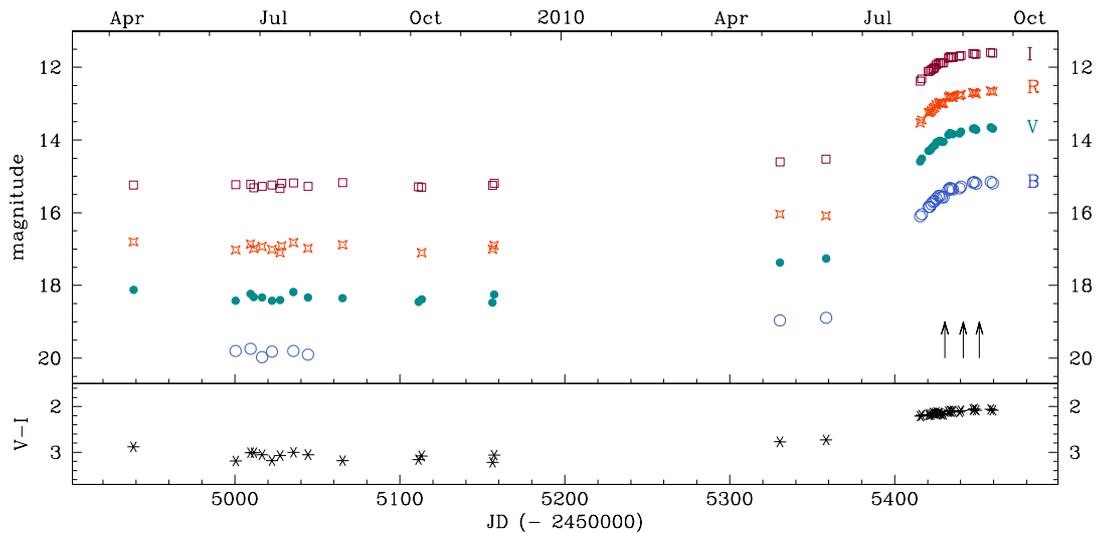}
      \caption{$BVRI$ light curves of HBC 722. Dates of spectroscopic observations are marked by arrows.}
         \label{Fig2}
  \end{figure*}
  
A pre-outburst spectrum of HBC 722 was obtained with the 1.3-m RC telescope of Skinakas Observatory on 2001 September 10.  
The observational procedure and data reduction process are described in Semkov (2003). 
Spectroscopic observations of HBC 722 during the period of increase in brightness were obtained with the 0.6-m telescope of Schiaparelli
Observatory in Varese (Italy), equipped with a multi-mode spectrograph (Echelle + single dispersion modes) and a SBIG ST10-XME CCD camera
(2184$\times$1472 array, 6.8~$\mu$m pixels).  
Table 3 provides a journal of spectroscopic observations.  
The spectra were treated for bias, dark, and flat fields in a standard way with IRAF following the recipes of Zwitter and Munari (2000).  
The spectra were calibrated into absolute fluxes by observations of some spectrophotometric standards preceding and following the exposures of HBC 722.

\begin{table}
\caption{Journal of spectroscopic observations}
\label{table:3}
\centering
\begin{tabular}{@{~~~}c@{~~~}c@{~~~}r@{~~~}c@{~~~}c@{~~~}}
\hline\hline
\noalign{\smallskip}
        Date  & $<$UT$>$  & exp. & disp.  & range\\
              &           & (sec)& (\AA/pix) & (\AA)\\
\noalign{\smallskip}
            \hline
            \noalign{\smallskip}
    2001-09-10 & 01:01  &  3600  &  1.04 & 5182-7265\\
    2010-08-21 & 22:06  &  3600  &  2.13 & 4400-7800\\
    2010-09-02 & 01:08  & 14400  &  2.13 & 4210-8850\\
    2010-09-11 & 21:12  & 14400  &  0.72 & 5670-7230\\
\hline                                   
\end{tabular}
\end{table}

\section{Results and discussion}

The $\it BVRI$ light-curves of HBC 722 during the period April 2009 - September 2010 are plotted in Fig.  2.  
Prior to the outburst (Apr.  - Nov. 2009), the star displayed only small amplitude variability.  
The observational data indicate that the outburst started sometime before May 2010, and reached its maximum value in September 2010.  
By comparing with brightness levels in 2009, we derive the following values for the outburst amplitude: $\Delta$$I$=3\fm7, $\Delta$$R$=4$\fm$3, $\Delta$$V$=4\fm7, and $\Delta$$B$=4\fm7.  
A corresponding increase in the infrared brightness was reported by Leoni et al. (2010), whose data for 2010 September 1, when compared
with 2MASS values provide the outburst amplitudes $\Delta$$J$=3$\fm$2, $\Delta$$H$=3\fm0, and $\Delta$$K$=2\fm8. 
Simultaneously with the increase in brightness the star color changes significantly, becoming appreciably bluer.  
An important result from the $BVRI$ photometry is that, while both $V$$-$$R$ and $R$$-$$I$ became bluer, at the same time $B$$-$$V$ did not change relative to quiescence.  
During the period of rise in brightness, the light curve of HBC 722 is similar to the light curves of the classical FUor object V1057 Cyg and FU Ori itself (Clarke et al. 2005).

Compared to quiescence, the spectrum of HBC changed significantly during outburst.  
The quiescent spectra of Cohen \& Kuhi (1979) and our one in Figure~4 are remarkably similar.  
While showing the Mg I triplet (5174 \AA), they both lack the Na I 5890, 5896, Li I 6707, and Ba II 5854, 6497 \AA\ absorption lines that are normally so prominent in outbursts (cf Fig.~5).  
The H$\beta$ and [OI] 6300 \AA\ lines reported in emission by Cohen \& Kuhi (1979) have disappeared in outburst, and H$\alpha$ remains in emission, but has a smaller equivalent width (probably the results of a much brighter background continuum). 
Figure~6 shows the evolution of the H$\alpha$ profile during the outburst phase, highlighting the appearance of a strong absorption component in the
spectrum for 2010 September 11.

\begin{figure}
   \centering
   \includegraphics[width=9cm]{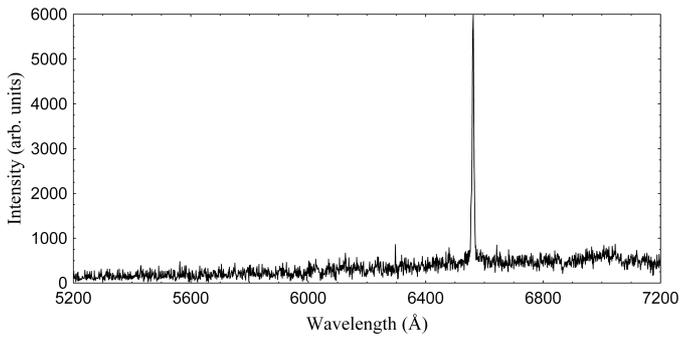}
   \caption{Pre-outburst spectrum of HBC 722 obtained on 2001 Sep 10}
    \end{figure}

\begin{figure}
   \centering
   \includegraphics[width=9cm]{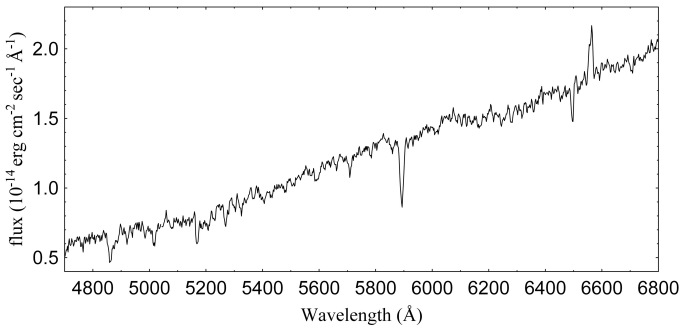}
   \caption{Outburst spectrum of HBC 722 obtained on 2010 Sep 2}
    \end{figure}    

\begin{figure}
   \centering 
   \includegraphics[width=9cm]{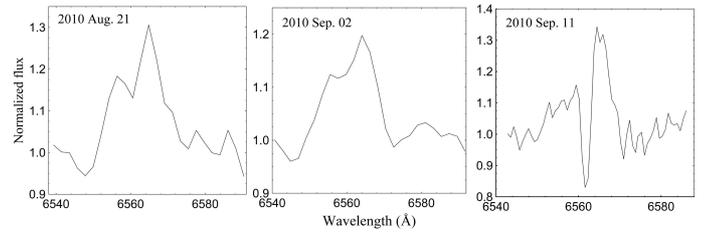}
   \caption{The profile of H$\alpha$ emission line extracted from the three
   2010 spectra of HBC 722 obtained during the increase in brightness (cf
   Table~3).}
    \end{figure}  

Photometric and spectroscopic data so far accumulated suggest that HBC 722 is indeed proceeding through a {\it bona fide} FUor outburst.  
The $\Delta$V=4\fm7 amplitude, the spectrum turning from emission-line dominated to one with prominent absorption lines of Li I 6707 and Ba II 6497, the
appearance of a strong NaI absorption, and the emergence of a reflection nebula around the star are all well-established characteristics of FUor
outbursts.  
The only discrepancy is the H$\alpha$ profile that remains dominated by the emission component even during the outburst. 
The profile of the H$\alpha$ line is highly variable during maximum light and the decrease in brightness, but for all known FUor stars is characteristic of a P Cyg profile dominated by absorption (Herbig et al. 2003). 
On the other hand, none of the classical FUors has spectral observations during the period of increase in brightness and we do not know exactly when the absorption engulfs the emission component.  
The outburst of HBC 722 differs from EXors eruptions in both duration and spectral appearance.  
However, a similarity with V1647 Ori, an object intermediate between FUors and EXors (Aspin \& Reipurth 2009), cannot yet be excluded.

Given the small number of known FUor objects, photometric and spectral studies of HBC 722 are of great interest.  
We encourage those interested in FUor and EXor variables to follow the star during the next few months.  
We plan ourselves to continue with our spectroscopic and photometric monitoring.

\begin{acknowledgements}
      This work was partly supported by grants DO 02-85, DO 02-273 and DO
      02-362 of the National Science Fund of the Ministry of Education,
      Youth and Science, Bulgaria.  The authors thank the Director of
      Skinakas Observatory Prof.  I.  Papamastorakis and Prof.  I. 
      Papadakis for the telescope time.  This research has made use of the
      NASA Astrophysics Data System.
\end{acknowledgements}

\end{document}